\documentclass[preprint,showpacs,preprintnumbers,amsmath,amssymb]{revtex4}

\usepackage{graphicx}
\usepackage{dcolumn}
\usepackage{bm}

\newcommand{\sfrac}[2]{{\scriptsize \frac{#1}{#2}}}
\def\d{\textrm{d}}
\def\comment#1{}

\def\Re{{\rm Re\,}}

\def\mbf#1{#1}

\begin{document}


\title{Superpositions of Probability Distributions}
\author{Petr Jizba}
\email{jizba@physik.fu-berlin.de} \altaffiliation[On leave
from\\]{~FNSPE, Czech Technical University, B\v{r}ehov\'{a} 7, 115
19
Praha 1, Czech Republic}
\author{Hagen Kleinert}%
\email{kleinert@physik.fu-berlin.de} \affiliation{ITP, Freie
Universit\"{a}t Berlin\\ Arnimallee 14 D-14195 Berlin, Germany}
\date{\today}

\begin{abstract}
Probability distributions which can be obtained from superpositions
of Gaussian distributions of different variances
$v= \sigma ^2$  play a favored role  in quantum theory and financial
markets. Such superpositions need not necessarily
obey the Chapman-Kolmogorov semigroup relation for
Markovian processes because they may introduce memory effects. We
derive the general form of the smearing distributions  in $v$
which do not destroy the semigroup property.
The smearing technique has two immediate
applications.  It permits simplifying the system of Kramers-Moyal
equations for smeared and unsmeared conditional probabilities, and
can be conveniently implemented in the path integral
calculus.  In many cases, the superposition of path integrals
can be evaluated much easier than the initial path integral.
Three simple examples are presented, and it is shown how the technique is extended
to quantum mechanics.
\end{abstract}

\maketitle

\section{Introduction}

Path integrals are a powerful tool in diverse areas of physics, both
computationally and conceptually. They often provide the easiest
route to derivation of perturbative expansions as well as an
excellent framework for nonperturbative analysis~\cite{PI}. One of
the key properties of path integrals in statistical physics is that
the related time evolution of the conditional probabilities
$P(x_b,t_b|x_a,t_a)$ fulfills the Chapman-Kolmogorov (C-K) equation
for continuous Markovian processes
\begin{equation}
P(x_b,
t_b|x_a,t_a)=\int \d x\, P(x_b, t_b|x, t) P(x, t |x_a,t_a).
 \label{@CK}\end{equation}
Conversely, any probability satisfying this equation possesses a
path integral representation, as shown by Kac and
Feynman~\cite{Kac,FH,feller}. Equation~(\ref{@CK}) also serves as a
basis for deriving a Fokker-Planck time evolution
equation~\cite{PI,Haba,Gardiner} for $P(x_b,t_b|x_a,t_a)$ from
either the stochastic differential equation obeyed by the variable
$x$ or the Hamiltonian driving the time evolution of
$P(x_b,t_b|x_a,t_a)$. Such equations are used to explain many
different physical phenomena, for example turbulence~\cite{FP} or
epitaxial growth~\cite{M-V}. In information theory they serve as a
tool for modeling various queueing processes~\cite{GH1}, while in
mathematical finance they are conveniently applied in theory of
option pricing for efficient markets~\cite{PI,jkh07,TIMX,MNM}.

A trivial property of $ P(x_b, t_b|x_a, t_a)$ satisfying the C-K
equation (\ref{@CK}) is the initial condition
\begin{equation}
P(x_b,
t_a|x_a,t_a)= \delta (x_b-x_a).
\label{@INC}\end{equation}
The right-hand side can be written as a scalar product of Dirac's
bra and ket states $\langle x_b|$ and $|x_a\rangle$ as
\begin{equation}
P(x_b,t_a|x_a,t_a)= \langle x_b|x_a\rangle.
\label{@BRAK}\end{equation}

In many practical applications one encounters conditional
probabilities formulated as a superposition of path integrals in
which the Hamiltonians $H$ are rescaled by a factor $v$, i.e.,
\begin{eqnarray}
\bar{P}(x_b,t_b|x_a,t_a) \ = \ 
 \int_{0}^{\infty} \!\!\d v \ \omega(v,t_{ba})
\int_{x(t_a) = x_a}^{x(t_b) = x_b} \!\!\!\!\!{\mathcal{D}} x
{\mathcal{D}} p\ e^{\int_{t_a}^{t_b} \d \tau \left[i p\dot{x} - v
H(p,x) \right]}, \label{1.1}
\end{eqnarray}
where $ \omega(v,t_{ba})$ is some positive, continuous and
normalizable smearing function of $v\geq0$ and $ t_{ba}\equiv t_b-t_a \geq0$.
For instance, probability distributions which can be obtained from
superpositions of Gaussian distributions of different volatilities
$v= \sigma ^2$ play an important role in financial
markets~\cite{PI,jkh07}. Such smearing distributions show up also in
nonperturbative approximations to quantum statistical partition
functions~\cite{FK}, in systems with time reparametrization
invariance~\cite{PI,Pol}, in polymer physics~\cite{PI,AKH}, in
superstatistics~\cite{1}, etc. Whenever smeared path integrals
fulfill the C-K equation, the Feynman-Kac formula ensures that such
superpositions can themselves be written as a path integral without
smearing
\begin{eqnarray}
\bar P(x_b,t_b|x_a,t_a)=
\int_{x(t_a) = x_a}^{x(t_b) = x_b}\!\!\! {\mathcal{D}} x
{\mathcal{D}} p\ e^{\int_{t_a}^{t_b} \d \tau \left[i p\dot{x} - \bar
H(p,x) \right]}, \label{1.1'}
\end{eqnarray}
with a new Hamiltonian $\bar H(p,x)$ given by
\begin{equation}
 e^{ -\int_{t_a}^{t_b} \d \tau \bar H(p,x) }=
\int_{0}^{\infty} \d v \ \omega(v,t_{ba})
 e^{ -\int_{t_a}^{t_b} \d \tau  v
H(p,x) }\,.
\label{1.1111}
\end{equation}
%
In general, the smeared path integral (\ref{1.1}) does not conserve the C-K equation.
Physically this implies that the superposition (\ref{1.1}) introduces
memory into the system. The aim of this note is to find conditions
for the smearing distributions where this is avoided.
Various physical consequences will be derived from this.

The paper is organized as follows. Section~\ref{SEc2a} starts with a
warm up example of smearing distributions for a Gaussian conditional
probability. In Section~\ref{SEc2} we derive the most general class
of continuous smearing distributions fulfilling C-K equation.
Section~\ref{SEc3} is devoted to a construction of the Hamiltonian
$\bar{H}$ and to a discussion of related issues. In
Section~\ref{SEc4} we show how the outlined path integral
representation can be physically interpreted in terms of two coupled
stochastic processes. Section~\ref{SEc5} discusses three specific
smeared systems without memory. In particular we show how the
explicit knowledge of $\bar{H}(p,x)$ can streamline practical
calculations of numerous path integrals. Various remarks and
generalizations are proposed in the concluding Section~\ref{SEc7}.
For reader's convenience we include two appendices where we perform
some finer mathematical manipulations needed in Section~\ref{SEc4}.

\section{Smearing of a Gaussian distribution} \label{SEc2a}

Our goal is to find the most general form of $\omega(v,t)$
fulfilling the C-K relation (\ref{@CK}). Let us first illustrate
what we want to achieve by smearing out a simple Gaussian system
whose Hamiltonian is $H=vp^2/2$ leading to a conditional probability
\begin{eqnarray}
P_v(x_b,t_b|x_a,t_a) &=& \int_{x(t_a) = x_a}^{x(t_b) = x_b} {\mathcal{D}} x
{\mathcal{D}} p\ e^{\int_{t_a}^{t_b} \d \tau \left[i p\dot{x} - v
p^2/2 \right]}\nonumber \\[1mm]
&=&
\frac{1}{\sqrt{2\pi v t_{ba}}}\ \! e^{-(x_b - x_a)^2/(2 v t_{ba})}\, ,
\label{1.1aa}
\end{eqnarray}
where $t_{ba}\equiv  t_b - t_a$. This obeys the
{\em Fokker-Planck equation\/}
\begin{equation}
\partial _{t_b} P_v(x_b,t_b|x_a,t_a)=
 \sfrac{v}{2}\ \!\partial _{x_b}^2  P_v(x_b,t_b|x_a,t_a).
\label{@CK0}\end{equation}
which can be solved explicitly, including the initial
condition (\ref{@INC}) with the help of the differential operator for
the momentum $\hat p\equiv -i\partial _x$ as
\begin{eqnarray}
P_v(x_b,t_b|x_a,t_a) &=&
 e^{- t_{ba}v\hat{p}^2/2}  \delta (x_b-x_a),
\label{1.1aa'}
\end{eqnarray}
or, using the Dirac  {\em bra} and {\em ket} states in
(\ref{@BRAK}), as
\begin{eqnarray}
P_v(x_b,t_b|x_a,t_a) &=&
 \langle x_b| e^{- t_{ba}v\hat{p}^{2}/2} |x_a\rangle.
\label{1.1aa''}
\end{eqnarray}
Due to the completeness relation $\int \d x\, | x\rangle\langle x|=1$,
this expression obviously satisfies the C-K equation (\ref{@CK}).

Let us now assume that $\omega(v,t)$ can be written as a Fourier
transform of the form [compare also Ref.~\cite{beck1}]:
\begin{eqnarray}
\omega(v,t_{ba}) = \int_{-i\infty}^{i\infty} \frac{\d \xi}{2\pi i} \
e^{\ \!\xi v - H_{\omega}(\xi/t_{ba})t_{ba}}\, ,
\label{@FOU}\end{eqnarray}
~\\
where $H_{\omega}(\xi)$ can be viewed as the Hamiltonian affiliated
with distribution $\omega$. Then the smeared transition
probability has the integral representation
\begin{eqnarray}
\mbox{\hspace{-6mm}}\bar P(x_b,t_b|x_a,t_a) \!\! &=& \! \!
\int_{-i\infty}^{i\infty} \frac{\d \xi}{2\pi i} \ \! \bigg\langle
\!x_b\bigg|\frac{e^{-H_{\omega}(\xi/t_{ba})t_{ba}}}{\xi - t_{ba}
\hat{p}^2/2} \bigg|x_a\!\bigg\rangle. \label{1.1ab}
\end{eqnarray}
Assuming that  $H_{\omega}(\xi)$ is a regular function which becomes
infinite on a large semicircle in the upper half plane we can use
the residue theorem to evaluate this as
\begin{eqnarray}
\mbox{\hspace{-4mm}}\bar P(x_b,t_b|x_a,t_a)  =  \langle x_b |
e^{-H_{\omega}(i \hat{p}^2/2)t_{ba}} |x_a \rangle \, .
\label{1.1ac}
\end{eqnarray}
\\
This can be written as a path integral
\begin{eqnarray}  
\bar P(x_b,t_b|x_a,t_a)\ = \ \int_{x(t_a) = x_a}^{x(t_b) = x_b}
{\mathcal{D}} x {\mathcal{D}} p\ e^{\int_{t_a}^{t_b} \d
\tau \left[i p\dot{x} -  H_{\omega}(p^2/2) \right]}.
\label{1.1a}
\end{eqnarray}
~\\
Thus the Hamiltonian $\bar H(p,x)$ of the smeared system in
Eq.~(\ref{1.1'}) is simply equal to $ H_{\omega}(p^2/2)$.

In the following we shall generalize this treatment
to non-Gaussian Hamiltonians in Eq.~(\ref{1.1aa}).
~\\[-.8em]

\section{Smearing of a general distribution} \label{SEc2}

We now embark on finding the most general smearing function
$\omega(v,t_{ab})$ to guarantee the C-K relation for the smeared
expression (\ref{1.1}). Replacing the probabilities in (\ref{@CK})
by (\ref{1.1}), we bring the right-hand side to the explicit form
\begin{eqnarray}
&&\mbox{\hspace{-11mm}}\int_{0}^{\infty} \d v' \ \omega(v',t')
\int_{0}^{\infty} \d v'' \
\omega(v'',t'') \ \int_{-\infty}^{\infty} \d x \ \int_{x(t_c) = x}^{x(t_b) = x_b}
{\mathcal{D}} x {\mathcal{D}} p \ e^{\int_{t_c}^{t_b} \d \tau \left(ip\dot{x} - v''
H \right)} \nonumber \\[2mm]
&&\mbox{\hspace{6cm}}\times \ \int_{x(t_a) =
x_a}^{x(t_c) = x} {\mathcal{D}} x {\mathcal{D}} p \ e^{\int_{t_a}^{t_c} \d \tau
\left(ip\dot{x} - v' H \right)} \nonumber \\[4mm]
&&\mbox{\hspace{-11mm}}= \ \frac{t}{2 t' t''} \int_{0}^{\infty}\!\!\! \d v \int_{-tv}^{tv}
\!\!\!\d \zeta \ \omega\left(\frac{\left(tv - \zeta\right)}{2t'}
,t'\right)\omega\left(\frac{\left(tv + \zeta\right)}{2t''}
,t''\right)\ \int_{x(t_a) = x_a}^{x(t_b) = x_b} {\mathcal{D}} x {\mathcal{D}} p \
e^{\int_{t_a}^{t_b} \d \tau \left(ip\dot{x} - v H \right)} ,
\label{1.3}
\end{eqnarray}
where $t' \equiv t_{ca}=t_c- t_a$ and  $t'' \equiv
t_{bc}= t_b - t_c$. In the second line we have used
the substitution $t''v'' + t'v' = tv$ and $t''v'' - t'v' = \zeta$.
Comparing the right-hand side of (\ref{1.3}) with the left-hand
side of (\ref{1.1}) expressed in the smeared form (\ref{1.1}), we obtain
an integral equation for the smearing function $\omega(v,t)$:
\begin{eqnarray}  \!\!
\int_{-tv}^{tv}\! \!\!\!\!\d \zeta \ \omega\!\left(\!\frac{\left(tv
- \zeta\right)}{2t'} ,t'\!\right)\!\omega\!\left(\!\frac{\left(tv +
\zeta\right)}{2t''} ,t''\!\right)\! = \frac{2t't''\omega(v,t)}{t} .
\label{1.4}
\end{eqnarray}
Setting $z
\equiv tv/t''$ and going over to an integration variable $z' = (tv +
\zeta)/2t''$,
this becomes
\begin{eqnarray}
\int_{0}^{z}\! d z' \
\omega(z',t'')\omega\!\left(\!\frac{t''}{t'}(z-z'),t'\!\right)=
\frac{t'}{t}\ \!\omega\left(\!\frac{t''}{t}z,t\!\right), \label{1.5}
\end{eqnarray}
or equivalently
\begin{eqnarray}
\int_{0}^{z}\! d z' \ \omega(z',t)\ \! a \ \!
\omega\!\!\left(\!a(z-z'),\frac{t}{a}\! \right)  = b\ \!
\omega\!\!\left(\!b z,\frac{t}{b}\! \right)\! ,
 \label{1.5b}
\end{eqnarray}
with positive real $a,b$ satisfying $1 + 1/a = 1/b$.
Since the left-hand side is a convolution integral, the solution
of this equation is found by a Laplace transformation.
Defining
%
%
\begin{eqnarray}
\tilde{\omega}(\xi,t) = \int_0^{\infty} \d z \ e^{-\xi z}
\omega(z,t), \;\;\;\;\; \Re \xi >0\, ,
\label{1.6}
\end{eqnarray}
we may reduce Eq.~(\ref{1.5b}) to the functional equation
\begin{eqnarray}
\tilde{\omega}(\xi,t)\
\!\tilde{\omega}\!\left(\!\frac{\xi}{a},\frac{t}{a}\! \right) =
\tilde{\omega}\!\left(\frac{\xi}{b},\frac{t}{b}\right)\, .
\label{1.7}
\end{eqnarray}
It should be stressed that due to assumed normalizability and
positivity of $\omega(v,t)$, the smearing distribution is always
Laplace transformable. The substitution $\alpha =1/a$ transforms
(\ref{1.7}) to
\begin{eqnarray}
\tilde{\omega}(\xi,t)\ \!\tilde{\omega}\!\left(\alpha \xi, \alpha
t\! \right) = \tilde{\omega}\!\left(\xi + \alpha\xi, t + \alpha
t\!\right)\, . \label{1.8}
\end{eqnarray}
By considering only real $\xi$, Eq.~(\ref{1.8}) can be solved by
method of iterations familiar from the theory of functional
equations~\cite{acz1}. Assume for a moment that $\alpha$ is a
positive integer, say $n$, then successive iterations of
Eq.~(\ref{1.8}) give
\begin{eqnarray}
\tilde{\omega}(n\xi,n t) =  [\tilde{\omega}(\xi,t)]^n\, .
\label{1.8b}
\end{eqnarray}
Now let $r = m/n$ be a positive rational number ($m$ and $n$
positive integers) and $\zeta$ and $\tau$ arbitrary
positive real numbers. Then, for $\xi = r\zeta = (m/n)\zeta$ and
$t = r\tau = (m/n)\tau$, we have $n\xi = m\zeta$ and $nt = m\tau$,
so that Eq.~(\ref{1.8b}) yields
\begin{eqnarray}
\tilde{\omega}(\xi,t) \ = \ \tilde{\omega}(r\zeta, r\tau) \ = \
[\tilde{\omega}(\zeta,\tau)]^r\, ,
\label{1.8c}
\end{eqnarray}
for all positive $\zeta$ and $\tau$ and all positive rationals $r$.
Assuming that $\tilde{\omega}$ is continuous, we may extend the
Eq.~(\ref{1.8c}) to all positive real $r$. It is then solved by
\begin{eqnarray}
\tilde{\omega}(\xi,t) \ &=& \ \tilde{\omega}(t \cdot \xi/t, t \cdot
1 )\nonumber \\ \ &=& \ [\tilde{\omega}(\xi/t,1)]^{t} \ \equiv \
[G(\xi/t)]^t , \;\;\;\; t> 0\, , \label{1.9c}
\end{eqnarray}
where $G(x)$ is any continuous function of $x$. The above derivation
is meaningless for $t=0$. In this case we must instead of
(\ref{1.9c}) consider
\begin{eqnarray}
\tilde{\omega}(\xi,0) \ = \ \tilde{\omega}(\xi \cdot 1, \xi \cdot 0)
\ = \ [\tilde{\omega}(1,0)]^{\xi} \ \equiv \ {\kappa}^{\xi}  \, .
\label{1.9d}
\end{eqnarray}
The constant $\kappa$ is determined by the initial value of the smearing
distribution $\omega(v,t)$. Thus Eq.~(\ref{1.9d}) implies that
\begin{eqnarray}
\lim_{t \rightarrow +0} \omega(v,t) &=& \theta(v + \log \kappa)
\delta(v + \log \kappa)\nonumber \\  &=& \delta^+(v + \log \kappa)\,
.\label{1.9e}
\end{eqnarray}
Note that Eq.~(\ref{1.9c}) implies positivity of
$\tilde{\omega}(\xi, t)$ for all $t$ and $\xi$, hence $G(x)$
must also be positive for all $x$. This allows us to write
\begin{eqnarray}
[G(\xi/t)]^{t} = e^{-F(\xi/t)t},
 \label{@ExpO}\end{eqnarray}
where $F(x)$ is some continuous function of $x$. The associated
inverse Laplace transform gives then the complete solution for
$\omega(v,t)$. Usually, the inverse Laplace transform is expressed
as as a complex Bromwich integral~\cite{BW} [compare (\ref{@FOU})]:
\begin{eqnarray}
\omega(v,t) = \frac{1}{2\pi i} \int_{\gamma - i\infty}^{\gamma + i
\infty} \d\xi \ e^{\xi v}\ \! \tilde  \omega (\xi,t)
\, ,\label{1.8e}
\end{eqnarray}
where the real constant $\gamma$ is such that it exceeds the real
part of all the singularities of $\tilde  \omega (\xi,t)$. For our
purpose it will be preferable to use, instead of (\ref{1.8e}), the
inversion formula due to Post~\cite{post}:
\begin{eqnarray}
\omega(v,t) = \lim_{k\rightarrow \infty} \frac{(-1)^k}{k!}
\left(\frac{k}{v}\right)^{\!\!k+1} \left.
{\partial_\xi^k\tilde{\omega}(\xi,t)}\right|_{\xi = k/v}\, .
\label{2.15}
\end{eqnarray}
For practical calculations this formula is  rarely used due to the
need to evaluate derivatives of arbitrary high orders. For our
purpose, however, it has the advantage that it shows that a real
Laplace transform $\tilde{\omega}(\xi,t)$ leads to a real smearing
function $ \omega (v,t_{ab})$. Moreover, one does not need to know
the pole structure of $\tilde{\omega}(\xi,t)$ in the complex
$\xi$-plane, which is in general not known.

The result (\ref{1.9c}) can be mildly generalized to the smearing of
composite Hamiltonians $vH(p,x) = vH_1(p,x) + H_2(p,x)$ as long as
$[\hat{H}_1,\hat{H}_2] = 0$. At the same time it should be stressed
that the entire approach fails for time-dependent Hamiltonians.

To end this section we note that if the smearing distribution
$\omega(v,t)$ would have included negative $v$-values, the integral
in Eq.~(\ref{1.5b}) would have been replaced with $\int_0^z \mapsto
\int_{\infty}^{-\infty}$. Then a two-sided Laplace transformation
would have  brought us to the same equations
(\ref{1.8c})--(\ref{1.9c}) as before, and the associated general
$\omega(v,t)$ would been recovered via the inverse of the two-sided
Laplace transform. By restricting ourselves to $v \geq0$, all
calculations become simpler.

\section{Explicit representation of $\bar H(p,x)$} \label{SEc3}

We now determine the Hamiltonian $\bar{H}(p,x)$ explicitly in the
general case. For this we must first make sure that the path
integral for the initial distribution in (\ref{1.1})
\begin{eqnarray} 
P_v(x_b,t_b|x_a,t_a) \ = \ \int_{x(t_a) = x_a}^{x(t_b) =
x_b} {\mathcal{D}} x {\mathcal{D}} p\ e^{\int_{t_a}^{t_b} \d
\tau \left[i p\dot{x} - v H(p,x) \right]}, \label{1.1aaa}
\end{eqnarray}
is properly defined. The classical Hamiltonian
$ H(p,x)$ must be set up in such a way that the Hamiltonian operator
driving the time evolution
\begin{eqnarray}  \!\!
\partial _{t}P_v(x,t|x_a,t_a)=
- v\hat{H}(\hat p,x) P_v(x,t|x_a,t_a)\, ,
\label{@TIMEEV}\end{eqnarray}
has all momentum operators $\hat p=-i\partial _x$ to the left of the
$x$-variables. Only then can one guarantee the probability
conservation law for $ P(x_b,t_b|x_a,t_a) $:
\begin{eqnarray}
\int \d x \ \! \partial _{t}P_v(x,t|x_a,t_a)\ = \ - \int \d x\ \!
v\hat H(\hat p,x) P_v(x,t|x_a,t_a). \label{@2}\end{eqnarray}
The right-hand side vanishes after a partial integration. The
relation between $\hat H(\hat p,x)$ and $ H( p,x)$ is explained in
Ref.~\cite{PI}.

With the help of Post's inversion formula~(\ref{2.15}) we
now rewrite the smeared conditional probability (\ref{1.1}) in the form
\begin{eqnarray}
\mbox{\hspace{-0mm}}\bar P(x_b,t_b;x_a,t_a) &=&
 \ \lim_{k\rightarrow \infty} \frac{(-1)^{k-1}}{(k-1)!}  \int_{0}^{\infty}\!\!\d \xi \ \xi^{k-1}
{\partial_\xi^k\tilde{\omega}(\xi,t_{ba})}
 \nonumber
\\[1mm]
&\times&  \int_{x(t_a) = x_a}^{x(t_b) = x_b} \!\!\!\!\!{\mathcal{D}}
x {\mathcal{D}} p\ e^{\int_{t_a}^{t_b} \d \tau \ \! (i p\dot{x} -  k
H/\xi)}. \label{3.1a}\end{eqnarray}
Inserting here Eq.~(\ref{1.6}), the integration over $\xi$ turns into
\begin{eqnarray}
\int_{0}^{\infty}\!\!\d v \ v^k \omega(v,t_{ba})
\int_{0}^{\infty}\d \xi
\ \xi^{k-1}
 e^{-v\xi} \
\! \int_{x(t_a)
= x_a}^{x(t_b) = x_b} \!\!{\mathcal{D}} x {\mathcal{D}}
p   ~
e^{\int_{t_a}^{t_b} \d \tau \
  [i p\dot{x}- k  H /\xi]
}.  \label{@REWR}
\end{eqnarray}
The path integral can be written in Dirac operator form
(\ref{1.1aa''}) as
\begin{equation}
 \langle x_b| e^{-k t_{ba}\,\hat H/\xi} |x_a\rangle,
\label{@3}\end{equation}
so that (\ref{@REWR}) becomes
\begin{eqnarray}&&
\int_{0}^{\infty} \d v \ \! v^k \omega(v,t_{ba}) \ \!\Big\langle x_b\Big|
\int_{0}^{\infty}\!\!\d \xi\hspace{1pt}  \xi^{k-1}
 e^{-v\xi}
  e^{-k t_{ba}\,\hat H/\xi}\Big|x_a \Big\rangle .
  \label{@REWR2}
\end{eqnarray}
The $\xi$-integral yields a Bessel function of the second type
\begin{eqnarray}
2\left({k  t_{ba}\hat H/y}\right)^{k/2}K_k\left(2
 \sqrt{ ky t_{ba}\hat H}\right).
\label{@4}\end{eqnarray}
Applying the limiting form of the Bessel function for large index
\begin{eqnarray}
K_{k}(\sqrt{k} x) \ \sim \ \frac{1}{2} \ \! k^{-k/2} \ \!\Gamma(k)
\left(\frac{x}{2}\right)^{\!\!-k} \! e^{-x^2/4}\, , \label{3.1bc}
\end{eqnarray}
we can cast (\ref{3.1a}) to
\begin{eqnarray}
\bar P(x_b,t_b|x_a,t_a)\ = \ \bigg\langle\! x_b\bigg|\int_{0}^{\infty} \d v
\,\omega(v,t_{ba})  e^{- vt_{ba}\hat H } \bigg|x_a\! \bigg\rangle.
\end{eqnarray}
We now use Eqs.~(\ref{1.6}) and (\ref{1.9c}) to rewrite this as
\begin{eqnarray}
\langle x_b|\tilde  \omega ( t_{ba}\hat H
,t_{ba})|x_a\rangle \
= \ \langle x_b| \tilde  \omega ( \hat H
,1)^{t_{ba}} |x_a\rangle. \label{@INI}\end{eqnarray}
With Eq.~(\ref{@ExpO}),
this becomes
\begin{eqnarray}
\bar P(x_b,t_b|x_a,t_a) = \langle x_b| e^{-t_{ba} F(\hat H)}
|x_a\rangle, \label{@FINI0}\end{eqnarray}
which can be expressed as a path integral
\begin{eqnarray}
\bar P(x_b,t_b|x_a,t_a)\ = \  \int_{x(t_a) =
x_a}^{x(t_b) = x_b}\! \!\!\!\!{\mathcal{D}} x {\mathcal{D}} p  \
e^{\int_{t_a}^{t_b} \d t [ i p\dot{x} -  F_{\rm cl}(H)]} .
\label{@FINI}
\end{eqnarray}
Here $F_{\rm cl}(H)$ denotes the classical function of the energy
which makes the path integral (\ref{@FINI}) equal to the operator
expression $F(\hat{H})$ in (\ref{@FINI0}). The construction of
(\ref{@FINI}) is highly nontrivial since one must ensure that the
path integral leads to the correct operator order in $\tilde \omega
( \hat H ,1)^{t_{ba}}$. The task is simple only if $\hat{H}$ depends
only on $\hat{p}$. Then the ordering problem disappears and $F_{\rm
cl}(H(p))= F(H(p))$.

There are other instances in which the operator ordering can be uniquely assigned.
In Appendix~B we discuss one such example.

A few observations are useful concerning he nature of $F(H)$. First, the
normalization condition
\begin{eqnarray}
1 = \int_{0}^{\infty} \d v \ \omega(v,t) = \tilde{\omega}(0,t) \, ,
\label{3.1c}
\end{eqnarray}
implies that $F(0) = 0$. Second, the necessary positivity of
$\omega(v,t)$ implies, via Post's formula, that all derivatives
$(-1)^k \partial^k \tilde{\omega}(\xi,t)/\partial \xi^k$ are
positive for large $\xi$ and $k$. Taking into account that
$\tilde{\omega}(x,t)>0$ we conclude that $\tilde{\omega}(\xi,t)$
must be decreasing and convex for large $\xi$ and any $t$.
Asymptotic decrease and convexity of $e^{-F(x/t)t}$  ensure that
$F(\hat{H})$ must be a monotonically increasing function of
$\hat{H}$ for large spectral values of $\hat{H}$. Third, the real
nature of $\tilde{\omega}(x,t)$ makes $F(\hat{H})$ a real function
of $\hat{H}$. This is in contrast to quantum-mechanical path
integrals where the smearing function $\omega(v,t)$ is not
necessarily real.

\section{Kramers-Moyal expansions} \label{SEc4}

Let us study the  implications of the above smearing procedure upon
the time evolution equations for the conditional probabilities
$P_v(x_b,t_b|x_a,t_a)$ and $\bar P(x_b,t_b|x_a,t_a)$ which have the
general form (\ref{@TIMEEV}). In statistical  physics these are
called Kramers-Moyal (K-M) equations~\cite{Gardiner,PI,vK}. The
negative time evolution operator $-v\hat H(\hat p,x)$ is  called
{\em Kramers-Moyal operators\/} $ {\mathbb{L}_v}(-\partial _x,x)$.
Thus Eq.~(\ref{@TIMEEV}) for $P_v(x_b,t_b|x_a,t_a)$ and an analogous
equation for $\bar P(x_b,t_b|x_a,t_a)$ are written as \comment{ We
now turn to a consideration of the stochastic implications of the
fact that both $P(x_b,t_b|x_a,t_a)$ and
\begin{eqnarray}
\mbox{\hspace{-4mm}}P_v(x_b,t_b|x_a,t_a) \equiv  \int_{x(t_a) =
x_a}^{x(t_b) = x_b} \!\!\!{\mathcal{D}} x {\mathcal{D}} p\
e^{\int_{t_a}^{t_b} \d \tau \left(i p\dot{x} - v H \right)} \ .
\label{4.1}
\end{eqnarray}
fulfill the  C-K equations (\ref{@CK}) obey a Kramers-Moyal (K-M)
equation~\cite{Gardiner,PI,vK}. For $P_v(x_b,t_b|x_a,t_a)$, this
reads
}%
\begin{eqnarray}
{\partial}_{ t_b} P_v(x_b,t_b|x_a,t_a) &=&
{\mathbb{L}}_{v}\ \! P_v(x_b,t_b|x_a,t_a)\, ,\label{4.2a}\\
{\partial}_{ t_b} \bar{P}(x_b,t_b|x_a,t_a) &=& {\bar{\,\mathbb{L}}}\
\bar{P} \ \!(x_b,t_b|x_a,t_a)\, .
 \label{4.2b}
\end{eqnarray}
The Kramers-Moyal operator ${\mathbb{L}}_{v}$ has the
expansion
\begin{eqnarray}
{\mathbb{L}}_{v}(-\partial _{x_b},x_b) =
\sum_{n=1}^{\infty}\left(-{\partial }_{x_{b}}\right)^{\! n} \
\!D_v^{(n)}(x_b,t_b) \, , \label{4.3}
\end{eqnarray}
whose coefficients $D^{(n)}_v(x,t)$
are equal to the moments of the short-time transition probabilities:
\begin{eqnarray}
\mbox{\hspace{-9mm}~~}D^{(n)}_v(x,t)\! \!&=&\! \!\frac{1}{n!}
\lim_{\tau\rightarrow 0} \frac{1}{\tau}\!\int_{-\infty}^{\infty}
\!\!\!\d y\ \!
 (y\!-\! x)^n P_v(y,t +
\tau|x,t).
\label{@4.4a0}\end{eqnarray}
Inserting Eq.~(\ref{@TIMEEV}), these can also be calculated
from the formula
\begin{eqnarray}
\mbox{\hspace{-8.mm}~~}D^{(n)}_v(x,t) \!\!&=&\! \!
\frac{1}{n!}\lim_{\tau \rightarrow 0}\frac{1}{\tau }
\int_{-\infty}^{\infty}\!\! \d y \ \! (y\!-\! x)^n \ \!\langle
y|e^{-v \hat{H}\tau} |x \rangle . \label{4.4b}
\end{eqnarray}
%
%
%

The same equation holds for $\bar P(x_b,t_b|x_a,t_a)$ with the
replacement $v\hat H\rightarrow\hat{\bar H}$,
${\mathbb{L}}_{v}\rightarrow{\bar{\,\mathbb{L}}}$, and
$D_v^{(n)}(x,t)\rightarrow{\bar D}^{(n)}(x,t)$.
\comment{(\ref{4.2a})--(\ref{4.2}) can be cast into yet another,
equivalent system of equations.}

Our smearing procedure can be recast in the language of K-M
equations. We show that the two equations (\ref{4.2a}) and
(\ref{4.2b}) can be replaced by an equivalent pair of K-M equations.
First we observe that Eq.~(\ref{1.5}) can be rewritten as
\begin{eqnarray}
\omega(z,t)  =  \int_{0}^{\infty} \d z' \ \!
P_\omega(z,t|z',t')\omega(z',t')\, , \label{4.7}
\end{eqnarray}
with the conditional probability
\begin{eqnarray}
\!\!\!P_\omega(z,t|z',t')
=\frac{t}{t-t'} \theta(t z \!- \!t' z') \
\!\omega\!\left(\frac{t z \!- \!t' z'}{t\!-\!t'},
t-t'\!\right)\!,\label{4.8a}
\end{eqnarray}
which satisfies the initial condition
\begin{eqnarray}
&&\mbox{\hspace{-16mm}}\lim_{\tau \rightarrow
0}P_\omega(z,t+\tau|z',t) = \delta^+(z-z')\, . \label{4.8}
\end{eqnarray}
%

Moreover, expressing $\omega(z,t)$ in Eq.~(\ref{4.7}) in terms of
$P_\omega(x_b,t_b|x_a,t_a)$ via (\ref{4.8a}), we find that
$P_\omega(x_b,t_b|x_a,t_a)$ satisfies a C-K equation
\begin{equation}
P_\omega(x_b,
t_b|x_a,t_a)\ = \ \int \d x \ \! P_\omega(x_b, t_b|x, t) P_\omega(x, t |x_a,t_a).
 \label{@CKo}\!\!\!\!\!\,\end{equation}
This implies that
$P_\omega(x_b,
t_b|x_a,t_a)$
obeys a K-M time evolution equation
\begin{eqnarray} 
\partial_{t_{ba}} P_\omega(v_b,t_b|v_a,t_a)\ = \
{\mathbb{L}}_\omega  P_\omega(v_b,t_b|v_a,t_a)\, ,
\label{4.11ba}
\end{eqnarray}
with  the K-M operator
\begin{eqnarray}
{\mathbb{L}}_\omega = \sum_{n=1}^{\infty}
\left(-\partial_v\right)^{\! n} K^{(n)}(v,t_{ba})\, . \label{4.11c}
\end{eqnarray}
The expansion coefficients are obtained by analogy with
(\ref{@4.4a0}) from the short-time limits
\begin{eqnarray}
&&\mbox{\hspace{-9mm}}K^{(n)}(v,t)
=  \lim_{\tau\rightarrow 0} \frac{1}{n!\tau}
\int_{-\infty}^{\infty} \d v' \ \! (v'-v)^n
P_\omega(v',t+\tau|v,t)\nonumber \\[2mm]
&&\mbox{\hspace{-4mm}}=  \lim_{\tau\rightarrow 0}
\frac{1}{n!\tau}\left(\frac{\tau}{t_{ba} + \tau}\right)^n \!\!
\int^{\infty}_{0} \d v' \ \! (v'-v)^n \omega(v',\tau)\, .
\label{4.9}
\end{eqnarray}

A crucial observation for the further development is that due to the
equality of the Kernel in the time evolution equation (\ref{4.11ba})
and (\ref{4.7}), the same K-M operator governs the time evolution of
$ \omega(v,t_{ba})$:
\begin{eqnarray}
&&\partial_{t_{ba}} \ \! \omega(v,t_{ba}) = {\mathbb{L}}_\omega
\ \! \omega(v,t_{ba})\, , \label{4.11aa}
\end{eqnarray}
In Appendix A we show that this equation can also be derived directly
from the K-M equations (\ref{4.2a}) and (\ref{4.2b}).

The smearing procedure of the path integral
is equivalent to replacing the pair of
K-M equations (\ref{4.2a}) and (\ref{4.2b}) by the pair
(\ref{4.2a}) and (\ref{4.11aa}). This separates
the dynamics of the smearing distribution from
the dynamics of the transition amplitude.
This may serve as a convenient starting
point, for instance in quantum optics~\cite{QO}, in
superstatistics~\cite{1}, or in numerous option pricing models (see,
e.g., Ref.~\cite{PI} and citations therein).

For applications it is useful to remember Pawula's theorem~\cite{pawula},
according to which the coefficients of the
expansions of K-M operators are either all nonzero or, if there is
a finite number of them, they can be nonzero only up to $n=2$.
This follows from the necessary positivity of the probabilities.
If one artificially truncates $D_v^{(n)}(x,t)$
in (\ref{4.3}) or $K^{(n)}(x,t)$ at some $n
\geqslant 3$, then the ensuing transition probabilities {\em always}
develop negative values, at least for sufficiently short times.
This is the basic reason why phenomenological models for
K-M operators go usually only up to $n=2$.

Consider such a truncated model. Then the K-M equations
(\ref{4.2a}) and (\ref{4.11aa}) reduce to the Fokker-Planck
equations
\begin{eqnarray}
\mbox{\hspace{-5mm}} \partial_t \ \! \omega(v,t) &=&
{\mathbb{L}}_\omega^{\rm FP} \ \! \omega(v,t)\, ,
\label{4.11b} \\[2mm]
\mbox{\hspace{-5mm}} \partial_{t_b} P_v(x_b,t_b|x_a,t_a) &=&
{\mathbb{L}}_v^{\rm FP} P_v(x_b,t_b|x_a,t_a)\, , \label{4.11a}
\end{eqnarray}
with
\begin{eqnarray}
&&\mbox{\hspace{-5mm}}{\mathbb{L}}_\omega^{\rm FP} = -
\partial_v \ \! K^{(1)}(v,t) + \partial^2_v\ \!
K^{(2)}(v,t)\! , \label{4.11ab} \\[2mm]
&&\mbox{\hspace{-5mm}}{\mathbb{L}}^{\rm FP}_v =- \partial_{x_b} \
\!D^{(1)}_v(x_b,t_b) + \partial^2_{x_b}\ \! D^{(2)}_v(x_b,t_b)
\! . \label{4.11}
\end{eqnarray}
Furthermore, Eqs.~(\ref{4.11b}) and (\ref{4.11a}) allow us to find
two coupled stochastic processes  described by the two coupled
It\={o} stochastic differential equations
\begin{eqnarray}
&&\mbox{\hspace{-5mm}} \d x_b = D^{(1)}_v(x_b,t_b) \ \!\d t_b +
\sqrt{2 D^{(2)}_v(x_b,t_b)} \ \! \d W_1\, ,
\label{4.13a} \\[3mm]
&&\mbox{\hspace{-5mm}} \d v = K^{(1)}(v,t_{ba}) \ \! \d t_{ba} +
\sqrt{2 K^{(2)}(v, t_{ba})} \ \! \d W_2\, .
\label{4.13}
\end{eqnarray}
Here $W_1(t_b)$ and $W_2(t_{ba})$ are Wiener processes,
i.e.,  Gaussian random walks.
%

\section{Simple examples} \label{SEc5}

To demonstrate usefulness of the superposition procedure we now discuss
some important classes of $G(x)$-functions
in Eq.~(\ref{1.9c}). \\[2mm]
i.) We start with the trivial choice
\begin{eqnarray}
G(x) = e^{- ax + b}\ \! , \;\;\; b \in {\mathbb{R}}; \; a \in
{\mathbb{R}}^+_{0}\, , \label{5.1}
\end{eqnarray}
which gives
\begin{eqnarray}
\tilde{\omega}(\zeta,t) = e^{- a\zeta + b t}\, ,
 \label{5.2}
\end{eqnarray}
and consequently
\begin{eqnarray}
\omega(v,t) = e^{bt} \delta(v-a)\, . \label{5.3}
\end{eqnarray}
By requiring that $\omega$ is normalized to $1$ for any $t$  we have
$b =0$, i.e., no smearing distribution. In this case the Hamiltonian
$\bar{H}(p,x) = a H(p,x)$. Note that $a$ is basically the averaged
value of $v$ over the $\delta$-function
distribution.\\[2mm]
ii.) A less trivial choice of $G(x)$ is
\begin{eqnarray}
G(x) = \left(\!\frac{a}{x+ b}\!\right)^{\!\! c} \ \! , \;\;\;  a
\in {\mathbb{R}}^+; \; b,
c \in {\mathbb{R}}^+_0\, , \label{1.8f}
\end{eqnarray}
which gives
\begin{eqnarray}
\tilde{\omega}(\zeta,t) = \left(\frac{a t}{\zeta + b
t}\right)^{\!\!c t} \, , \label{1.10}
\end{eqnarray}
leading thus to
\begin{eqnarray}
\omega(v,t) = \frac{1}{\Gamma(ct)}(a t)^{c t} e^{-btv}
v^{ct-1} \, .
\label{1.11}\\[-2mm] \nonumber
\end{eqnarray}
Further restriction on the coefficients is obtained by requiring
normalizability of $\omega(v,t)$. The normalization condition $F(0)= 0$
can be fulfilled in two ways. Either we set $c=0$, in which case
$\omega(v,t) = \delta(v)$ and $\bar{H}(p,x) = 0$, or we assume that
$a=b$. In the latter case
\begin{eqnarray}
\omega(v,t) =  \frac{(bt)^{ct} v^{ct -1}}{\Gamma(ct)} \ \! e^{-bt v}
\, , \label{1.13}
\end{eqnarray}
i.e., it corresponds to the {\em Gamma}
distribution~\cite{feller,jkh07} $f_{bt, ct}(v)$ which is of a
particular importance in financial data
analysis~\cite{PI,Heston1,jkh07} and superstatistics~\cite{beck2}.
In the special case when $c = d b = d a$ and $b \rightarrow \infty$
we obtain that $\omega(v,t) = \delta(v-d)$. The Hamiltonian
$\bar{H}$ associated with the distribution (\ref{1.13}) reads
\begin{eqnarray}
\bar{H}(p,x) = \bar{v}b \ \! \left[\log\!\left(\!\frac{H(p,x)}{b} +
1\right)\right]_{\rm cl}, \label{5.13}
\end{eqnarray}
where $\bar{v} = c/b$ is the mean of $\omega(v,t)$. In particular,
for $H = {\mbf{p}}^2/2$ we obtain the smearing relation
between  path integrals
\begin{eqnarray}
&&\mbox{\hspace{-10mm}}\int_{{\mbf{x}}(t_{a}) =
{\mbf{x}}_a}^{{\mbf{x}}(t_{b}) = {\mbf{x}}_b} \!
 {\mathcal{D}} {\mbf{x}} {\mathcal{D}}
{\mbf{p}} \ e^{\int_{t_{a}}^{t_{b}} \d \tau \left[\! i
{\mbf{p}}\dot{\mbf{x}} - \bar{v}b\log\left({\mbf{p}}^2/2b \ + \
1\right)\right]}\nonumber \\
&&\!\!\!\!\!\!\!\!\!\!\!\!\!\!\!\!=\int_0^{\infty}\! \d v \,
 \omega (v,t_{ba})\! \int_{{\mbf{x}}(t_{a}) =
{\mbf{x}}_a}^{{\mbf{x}}(t_{b}) = {\mbf{x}}_b} \!\!\!\!
 {\mathcal{D}} {\mbf{x}} {\mathcal{D}}
{\mbf{p}} \ e^{\int_{t_{a}}^{t_{b}} \d \tau (i
{\mbf{p}}\dot{\mbf{x}} - v{\mbf{p}}^2/2)}\! .
\label{@THED}\end{eqnarray}
Inserting for the path integral  on the right-hand side
the result (\ref{1.3}), the integral over $v$  reads explicitly
\begin{eqnarray}  \!\!
 \int_{0}^{\infty} \!\!\!\! \d v \ \!
\frac{(bt_{ba})^{ct_{ba}} v^{ct_{ba} -1}}{\Gamma(ct_{ba})} \ \!
\sqrt{\frac{1}{2 \pi t_{ba} v}} \ \! e^{-bt_{ba} v}
e^{-  {{\mbf{x}}_{ba}^2}/{2vt_{ba}}}\! ,
\end{eqnarray}
and yields
\begin{eqnarray}  \!\!\! \!
K_{1/2 - c
t_{ba}}(2\sqrt{2b} \ \!|{\mbf{x}}_{ba}| )\ \!
\frac{t_{ba}^{-3/2}}{\sqrt{\pi}\ \! \Gamma(c t_{ba})}\left(\!\frac{2
\sqrt{2b} \ \!t_{ba}}{|{\mbf{x}}_{ba}|}\!\right)^{\!\! 1/2 -
ct_{ba}}\!\!\!. \label{6.35}
\end{eqnarray}
In Fourier space, the superposition (\ref{@THED}) leads to a Tsallis
distribution~\cite{jkh07,1,beck2}.

Another interesting consequence arises when we consider the ensuing
It\={o} stochastic equations. To this end we use the Hamiltonian
from Refs.~\cite{PI,jkh07} which has the form ${\mbf p}^2/2 + i{\mbf
p}(r/v -1/2)$, with $r$ being a constant. The corresponding drift
and diffusion coefficients $D_{v}^{(1)}$ and $D_{v}^{(2)}$ are
easily calculable giving
\begin{eqnarray}
\mbox{\hspace{-4mm}}D_{v}^{(1)}(x,t_{b}) \ = \  \left(r
-\frac{v}{2}\right),\;\;\  D_{v}^{(2)}(x,t_{b}) \ = \ \frac{v}{2} .
\end{eqnarray}
The coefficients $K^{(n)}$ are
\begin{eqnarray}
&&K^{(1)}(v,t_{ba}) \ = \ \frac{1}{t_{ba}}\left(\frac{c}{b} -
v\right) \ = \ \frac{1}{t_{ba}}
\left(\bar{v} - v\right) \, , \nonumber \\[2mm]
&&K^{(n)}(v,t_{ba}) \ = \ \frac{1}{t_{ba}^n} \frac{c}{n b^n} \, ,
\;\;\;\;\; n\geq 2\, . \label{5.14a}
\end{eqnarray}
Hence we find the two coupled  It\={o} equations (\ref{4.13a}) and
(\ref{4.13}) in the form
\begin{eqnarray}
&&\mbox{\hspace{-5mm}} \d x_{b} \ = \ \left(r -\frac{v}{2}\right) \d
t_{b} \ +
\sqrt{v} \ \! \d W_1\, , \nonumber \\[3mm]
&&\mbox{\hspace{-5mm}} \d v \ =  \ \frac{1}{t_{ba}}\left(\bar{v} -
v\right)\ \! \d t_{ba} + \frac{1}{t_{ba}} \sqrt{ \frac{\bar{v}}{b}}
 \ \d W_2\, . \label{6.36}
\end{eqnarray}
One may now view $x_b$ as a logarithm of a stock price $S$, and $v$
and $r$ as the corresponding variance and drift. If, in addition, we
replace for large $t_{ab}$ the quantity  $\sqrt{\bar{v}}$ with
$\sqrt{{v}}$, the systems (\ref{6.36}) reduces to
\begin{eqnarray}
&&\mbox{\hspace{-5mm}} \d S \ = \ r S \ \!\d t_{\!b} \ +
\sqrt{v} S\ \! \d W_1\, , \nonumber \\[3mm]
&&\mbox{\hspace{-5mm}} \d v \ =  \ \gamma\left(\bar{v} - v\right)\
\! \d t_{ba} + \varepsilon \sqrt{ {{v}}}
 \ \d W_2\, ,
\label{6.36b}
\end{eqnarray}
where $\gamma = 1/t_{ba}$ and $\varepsilon = 1/(b t_{ba})$. The
system of equations (\ref{6.36b}) constitute Heston's stochastic
volatility model~\cite{Heston1}. The parameters $\bar{v}$, $\gamma$
and $\varepsilon$ can be then interpreted as the long-time average,
the drift of the variance and the volatility of the variance,
respectively. Heston's model may be used in quantitative finance to
evaluate, for instance, the price of options (for further reference
on this model see, e.g., Ref.~\cite{PI} and citations therein).\\[2mm]
iii.)  As a third example we consider
\begin{eqnarray}
G(x) = e^{- a\sqrt{x }}\ \! , \;\;\;  a \in {\mathbb{R}}^+.
\label{5.14}
\end{eqnarray}
This implies
\begin{eqnarray}
\tilde{\omega}(\zeta,t) = e^{-a\sqrt{t\ \! \zeta}}\, ,\label{5.15}
\end{eqnarray}
and
\begin{eqnarray}
\omega(v,t) = \frac{a\ \!
e^{-a^2\!\frac{t}{4v}}}{2\sqrt{\pi}\sqrt{\frac{ v^3}{t}}}\, .
\label{5.16}
\end{eqnarray}
Image function $\tilde{\omega}(\zeta,t)$ already fulfils the
normalized condition $F(0) = 0$. In the literature, (\ref{5.16}) is
known as the {\em Weibull} distribution of order $1$ (e.g.,
Refs.\cite{feller,weibull}) $\omega(v,t) = w(v;1, a^2t/2)$.
The smearing distribution (\ref{5.16}) has also important
applications in the so called inverse $\chi^2$
superstatistics~\cite{beck2}. Because integral over
$v^{\alpha}\omega(v,t)$ does not exist for $\alpha > 1/2$, the
distribution (\ref{5.16}) does not have moments. The Hamiltonian
$\bar{H}(p,x) = a[\sqrt{H(p,x)}]_{cl}$. In particular, this implies
an identity \cite{remark}
\begin{eqnarray}
&&\mbox{\hspace{-5mm}}\int_{0}^{\infty}\!\!\!\!\!\d v \ \!
w\!\left(\! v;1, \frac{t}{2}\!\right)\! \int_{{\bf{x}}(t_a) =
{\bf{x}}_a}^{{\bf{x}}(t_b) = {\bf{x}}_b}\!\!\! {\mathcal{D}}{\bf{x}}
{\mathcal{D}}{\bf{p}}\ e^{\int_{t_a}^{t_b} \d \tau
\left[i {\bf{p}}\dot{\bf{x}} - v ({\bf{p}}^2 c^2 + m^2 c^4) \right]}\nonumber \\[2mm]
&&\mbox{\hspace{3mm}} = \ \int_{{\bf{x}}(t_a) =
{\bf{x}}_a}^{{\bf{x}}(t_b) = {\bf{x}}_b} \!\!\!{\mathcal{D}}{\bf{x}}
{\mathcal{D}} {\bf{p}}\ e^{\int_{t_a}^{t_b} \d \tau \left(i
{\bf{p}}\dot{\bf{x}} - c\sqrt{{\bf{p}}^2 + m^2 c^2} \right)}.
\label{5.17}
\end{eqnarray}
The right-hand side represents the path integral for free
relativistic particle in the Newton-Wigner
representation~\cite{NW,HK}. 
Previously, this has been evaluated by group path
integration~\cite{prugovecki}. With our smearing method, we can
obtain the same result much faster by a direct calculation of the
left-hand-side of (\ref{5.17}). Due the the quadratic nature of the
Hamiltonian we obtain immediately in $D$ space dimensions:
\begin{eqnarray}
&&\mbox{\hspace{-4mm}}\int_{{\bf{x}}(t_a) =
{\bf{x}}_a}^{{\bf{x}}(t_b) = {\bf{x}}_b} \!\!
 {\mathcal{D}} {\bf{x}} {\mathcal{D}}
{\bf{p}}\ e^{\int_{t_a}^{t_b} \d \tau
\left(i {\bf{p}}\dot{\bf{x}} -
c\sqrt{{\bf{p}}^2 + m^2 c^2} \right)}\nonumber \\[1mm]
&&\mbox{\hspace{-4mm}}= \ \int_{0}^{\infty}\d v \ \! \frac{
e^{-\!\frac{t_{ba}}{4v}}}{2\sqrt{\pi}\sqrt{\frac{ v^3}{t_{ba}}}}\ \!
e^{-v m^2c^4t_{ba} -
\frac{{\bf{x}}_{ba}^2}{4 v c^2 t_{ba}}}\!\left(\frac{1}{4vc^2 \pi t_{ba}}\right)^{\!\!D/2}
\nonumber \\[1mm]
&&\mbox{\hspace{-4mm}}=\ 2 c t_{ba }\left( \frac{m \gamma}{2\pi
t_{ba}} \right)^{\!\!\frac{D+1}{2}} \!\!
K_{\frac{D+1}{2}}\left(\frac{mc^2 t_{ba}}{\gamma}\right)\! ,
\label{5.18}
\end{eqnarray}
with $\gamma = (1 + {\bf{x}}_{ba}^2/c^2 t_{ba}^2)^{-1/2}$,
${\bf{x}}_{ba} \equiv {\bf{x}}_b - {\bf{x}}_a$.
The result
in
Eq.~(\ref{5.18}) agrees, of course,
with those of earlier authors~\cite{prugovecki}.

If Eq.~(\ref{5.18}) is taken to the limit $m=0$, it
reads
\begin{eqnarray}
\int_{{\bf{x}}(t_a) =
{\bf{x}}_a}^{{\bf{x}}(t_b) = {\bf{x}}_b}  {\mathcal{D}} {\bf{x}}
{\mathcal{D}} {\bf{p}}\ e^{\int_{t_a}^{t_b} \d \tau \left(i
{\bf{p}}\dot{\bf{x}} -
c|{\bf{p}|} \right)}\
= \ c t \ \!
\Gamma\!\left(\!\mbox{$\frac{D+1}{2}$}\!\right)\! \left(\!
\frac{\gamma^2}{\pi t^2 c^2} \!\right)^{\!\! (D+1)/2}\!
.\label{5.19}
\end{eqnarray}
This is of a particular importance in econophysics as it can
be directly related to a L\'{e}vy noise distribution of order $1$,
see, e.g. Ref.~\cite{PI}.

\section{Conclusions and outlooks} \label{SEc7}

In this paper we have identified the most general class of
continuous smearing distributions $\omega(v,t)$ defined on
$\mathbb{R}^+\!\times \mathbb{R}^+$ for which superpositions of
Marcovian processes remain Markovian. This insight was used to
rephrase the original problem of computing a complicated conditional
probability $P(x_b,t_b|x_a,t_a)$ as a problem solving two coupled
Kramers-Moyal equations, one for a simple basic distribution, and
the other for the smearing distribution $\omega(v,t)$. If the
associated Kramers-Moyal expansions are truncate after $n=2$, they
become Fokker-Planck equations whose respective sample paths follow
two coupled stochastic equations \`{a} la It\={o}. As an application
of this decomposition procedure we have demonstrated how the
log-normal fluctuations of a stock price with a Gamma smearing
distribution of the variances lead to the celebrated Heston model of
stochastic volatility.

As a second application we have shown that the ensuing relation
between smeared path integrals and non-smeared ones permits one to
solve exactly (and often quite fast) relatively large classes of
path integrals. In this connection we have discussed in some detail
two important situations: path integrals arising in the framework of
generalized statistics of Tsallis~\cite{tsallis1} and a world-line
representation of the euclidean Newton-Wigner propagator. The latter situation can
easily accommodate relativistic systems with gauge potentials via
minimal coupling. Useful applications are expected in the field of
world-line representations of effective actions~\cite{Pol,Schubert}.

For simplicity we have considered smearing distributions for path
integrals with only bosonic degrees of freedom. There is no problem
in extending our procedure to path integrals with Grassmann (i.e.,
fermionic) variables, and to more general initial functional
integrals. Such extensions may be  useful in polymer physics, in
particular in theory of self-avoiding chains, and in their
field-theoretic treatments~\cite{PI,Vilgis}.

Our procedure can also be applied to quantum mechanics with only
small modifications. There the C-K relation is the composition law
for time translation amplitudes reflecting the semigroup property of
the evolution operator $e^{-i\int dt\,H}$. The smearing functions
acts then upon probability amplitudes rather than conditional
probabilities, in which case they may be complex rather than
positive.

When the manuscript was finished we have became aware of
Refs.~\cite{beck1,beck2}, devoted to superstatistics. There exists a
conceptual overlap between the present paper and the above mentioned
ones. In fact, practitioners in superstatistics can quickly benefit
from our paper by substituting for our $v$ their intensive parameter
$\beta$ and for our smearing distribution $\omega(v,t)$ their
probability density $f(\beta)$. In superstatistical setting one
could also call our smearing procedure as ``the superstatistical
average" (though our distributions are generally time dependent). In
any case, it seems to us that important things are worth repeating
several times using different framework and different words. For a
more detailed discussion of the superstatistics paradigm, the reader
is referred to Refs.\cite{1,beck1,beck2}.


\begin{acknowledgments}
We acknowledge discussions with Z.~Haba and P.~Haener. Work is
partially supported by the Ministry of Education of the Czech
Republic (research plan no. MSM 6840770039), by the Deutsche
Forschungsgemeinschaft under grant Kl256/47.

\end{acknowledgments}

\section*{Appendix~A}\label{Ap1}

It is instructive to check that Eq. (\ref{4.11a}) can be derived
directly from the original K-M equations (\ref{4.2a}) and
(\ref{4.2b}) for $P_v(x_b,t_b|x_a,t_a)$ and $\bar
P(x_b,t_b|x_a,t_a)$. This provides an important cross-check for the
complete equivalence between both systems of K-M equations. To this
end, we multiply (\ref{4.2a}) by $\omega(v,t_{ba})$ and integrate
over $v$. This leads to
\begin{eqnarray}
&&\partial_{t_b} \bar{P}(x_b,t_b|x_a,t_a)\ - \
\int_0^{\infty} \d v \ \! P_v(x_b,t_b|x_a,t_a) \ \!\partial_{t_{ba}}
\ \!\omega(v,t_{ba}) \nonumber \\[2mm]
&&\mbox{\hspace{2.5cm}} = \ \int_{0}^{\infty} \d v \,\omega(v, t_{ba})
{\mathbb{L}}_{v} P_{v}(x_b,t_b|x_a,t_a). \label{@LABEL}
\end{eqnarray}
The first line has been rewritten using the chain rule for
derivatives, while in the second line we have used the fact that
$P_v(x_b,t_b|x_a,t_a)$ fulfills the K-M equation (\ref{4.2a}). We
now insert in the second line a trivial unit integral $\int
_0^\infty \d v'\, \omega (v', \tau)=1$, where $ \tau $ is
a positive
infinitesimal time. Then we subtract and add a term
$\bar{\,\mathbb{L}}\ \! \bar P(x_b,t_b|x_a,t_a)$, i.e., the
right-hand side of the K-M equation (\ref{4.2b}). Thus the second
line in (\ref{@LABEL}) becomes
\begin{eqnarray}
&&\mbox{\hspace{-0.4cm}} \int_{0}^{\infty}\!\! \d v
\!\int_{0}^{\infty}\!\! \d v' \omega(v, t_{ba}) \omega(v', \tau )
[{\mathbb{L}}_{v} - {\mathbb{L}}_{v'}]P_{v}(x_b,t_b|x_a,t_a)
\nonumber \\[2mm]&&~~~~\ + \ \bar{\,\,\mathbb{L}}\ \!\bar
P(x_b,t_b|x_a,t_a)\, . \label{A.0}
\end{eqnarray}
%
Eq. (\ref{A.0}) is a convenient short-hand version of more involved
form where the $\tau \rightarrow 0$ limit acts both on
$\omega(v,\tau)$ and ${\mathbb{L}}_{v}$. For instance,
\begin{eqnarray}
\ldots \int_{0}^{\infty}\!  \d v'  \omega(v', \tau )
{\mathbb{L}}_{v'}\ldots  \, ,
\end{eqnarray}
actually means
\begin{eqnarray}
&&\mbox{\hspace{-8mm}}\ldots
\sum_{k=1}^{\infty}\frac{1}{\tau k!} \!\int_{0}^{\infty}\!\! \d v'
\omega(v', \tau )
(-\partial_{x_b})^k \!\int_{-\infty}^{\infty} \!\!\d y \ \!(y- x_b)^k \ \! P_{v'}(y, t_b + \tau, x_b,t_b)\,
\ldots\, ,
\end{eqnarray}
and similarly for the ${\mathbb{L}}_{v}$-term.  With this in mind we
now Taylor-expand ${\,\mathbb{L}}_{v'}$ around $v$ and obtain
\begin{eqnarray}
&& - \sum_{n=1}^{\infty}
\frac{1}{n!}\int_{0}^{\infty} \d v \int_{0}^{\infty} \d v'\ \!
\omega(v, t_{ba}) \omega(v',  \tau ) (v'-v)^n\
\!\left[\partial^n_{v} \ \! {\mathbb{L}}_{v}\right] \ \!
P_{v}(x_b,t_b|x_a,t_a)  \nonumber \\[2mm] 
&&\mbox{\hspace{3cm}} + \ \bar{\,\,\mathbb{L}}\ \!
\bar P(x_b,t_b|x_a,t_a)\, . \label{A.0b}
\label{@LONG}\end{eqnarray}
To compute the derivatives $\partial^n_{v} \ \! {\mathbb{L}}_{v}$ we
use the representation (\ref{1.1aa'}) for a more general unsmeared
Hamiltonian operator $H(\hat p,x)$:
\begin{eqnarray}
P_{v}(x_b,t_b |x_a,t_a)  &=&
    e^{-vt_{ba} \hat{H}(\hat p_b,x_b)} \delta(x_b - x_a)\ = \
 e^{-vt_{ba} [\hat{H}^\dag(\hat{p}_a,x_a)]^*} \delta(x_b - x_a)
    \, ,
\end{eqnarray}
from which directly follows that\\[5mm]
\begin{eqnarray}
\partial_{v}^n \ \! P_{v}(y,t_b + \tau
|x_b,t_b)
&=& \ (- \tau )^{n}\hat{H}^ n(\hat p,y)
e^{-v\tau \hat{H}(\hat p,y)} \delta(y - x_b)\nonumber \\[2mm]
&=& \ (- \tau )^{n}[(\hat{H}^\dag)^
n(\hat{p}_b,x_b)]^* e^{-v\tau [\hat{H}^\dag(\hat{p}_b,x_b)]^*}
\delta(y - x_b) \, .
\end{eqnarray}\\[-3mm]
%
%
Appearance of the term $[\hat{H}^\dag(\hat{p},x)]^*$
in the second lines is a
consequence of the identity $\langle x_b|
\hat{A}(\hat{p},\hat{x})|x_a \rangle = \langle x_a|
\hat{A}^\dag(\hat{p},\hat{x})|x_b \rangle^*$ which is valid for any
operator $\hat{A}$. We can now simplify (\ref{A.0b}) with some
algebra involving product rules of derivatives and
$\delta$-functions.
Abbreviating $
\hat{H}(\hat{p},x)$
by $\hat{H}_x$,
this gives
\begin{eqnarray}
&&- \sum_{n=1}^{\infty} \frac{1}{n!\tau}
\left(\frac{\tau}{t_{ba} + \tau}\right)^{\!\!n}
\int_{0}^{\infty}\!\! \d v \int_{0}^{\infty}\!\! \d v'\ \! \omega(v,
t_{ba}) \omega(v', \tau ) (v'-v)^n \partial_{v}^n P_{v}(x_b,t_b +
\tau|x_a,t_a)\nonumber \\[2mm] 
&&\mbox{\hspace{3cm}} + \bar{\,\,\mathbb{L}}\ \! \bar{P}(x_b,t_b|x_a,t_a)\,
.\label{A.1}
\end{eqnarray}
In deriving the latter we have utilized the identity
\begin{eqnarray}
\int \d y \ \! (y-x_b)^k
[(\hat{H}^{\dag}_{y})^n]^* e^{-v\tau
(\hat{H}^{\dag}_y)^*} \ \!\delta(y - x_b) \
 = \ \int \d y \ \! \delta(y - x_b) \hat{H}_{y}^n
e^{-v\tau \hat{H}_y} (y-x_b)^k \, ,
\label{@YVAR}\end{eqnarray}
and the fact that
\begin{eqnarray}
\hat{H}^n_y e^{-\tau v \hat{H}_y} =
\left(\frac{-\partial_{\tau}}{v}\right)^{\!\!n} e^{-\tau v
\hat{H}_y}\, .
\end{eqnarray}
Extending for a moment the sum over $n$
in (\ref{@LONG}) to start from $n=0$,
the
relevant
terms entering the calculation of $\left[\partial^n_{v} \ \!
{\mathbb{L}}_{v}\right] \ \! P_{v}(x_b,t_b|x_a,t_a) $  in Eq. (\ref{A.1}) have the form
\begin{eqnarray}
&&\mbox{\hspace{-5mm}}\sum_{k=0}^{\infty}
\frac{(-\partial_{x_b})^k}{k!} \ \!\delta(y-x_b) e^{-\tau v H_{y}}\!
(y- x_b)^k e^{-t_{ba}v H_{x_b} }\delta(x_b-x_a)
\nonumber \\[1mm]
&&\mbox{\hspace{-5mm}}= \sum_{k=0}^{\infty}
\frac{(-\partial_{x_b})^k}{k!} \ \!\delta(y-x_b)[y(-i\tau) -y]^k
e^{-t_{ba}v H_y
}\delta(y-x_a)\nonumber \\[1mm]
&&\mbox{\hspace{-5mm}} = \ e^{[y- y(-i\tau)]\partial_{x_b}}\ \!
\delta(y-x_b)\ \! e^{-t_{ba}v H_y }\delta(y-x_a)\, .
\end{eqnarray}
We now apply the shift operator $e^{[y- y(-i\tau)]\partial_{x_b}}$
to
the $x_b$-variable in the $\delta$-function and cast the
resulting $\delta(x_b - y(-i\tau))$ into Dirac's bra-ket form
\begin{eqnarray}
\delta(x_b - y(-i\tau)) = \langle x_b | y(-i\tau)\rangle = \langle
x_b| e^{-\tau v \hat{H}} | y \rangle\, .
\end{eqnarray}
If we now  perform the integral over $y$
in (\ref{@YVAR}) and use the fact that
\begin{eqnarray}
\left(\frac{-\partial_{\tau}}{v} \right)^{\!\! n}\langle x_b|e^{-v(\tau
+ t_{ba}) \hat{H}}|x_a\rangle  \
= \  \left(\frac{-\partial_{v}}{t_{ba} +
\tau} \right)^{\!\! n}\langle x_b|  e^{-v(\tau + t_{ba})
\hat{H}}|x_a \rangle\, , \nonumber
\end{eqnarray}
we obtain the announced result (\ref{A.1}). We now use the
definition
of the moments
in the second line of Eq.~(\ref{4.9}),
and
the fact that the surface terms in the partial
integrations over $v'$
vanish due to
the positivity of the real part of the
Hamiltonian spectrum,
to rewrite the first term in
(\ref{A.1})
as
\begin{eqnarray}
&&\!\!\!\!\!\! -\! \sum_{n=1}^{\infty}
\int_{0}^{\infty}\!\!\! \d v  \! \left(-\partial_v \right)^n \!
\left[\!K^{(n)}({v},t_{ba}) \omega(v, t_{ba}) \right] \! \!
P_{v}(x_b,t_b|x_a,t_a)
\! . \nonumber
\end{eqnarray}
Remembering Eq.~(\ref{4.11c}), this can be written as
\begin{equation}
 -\int \d v\,{P}_v(x_b,t_b|x_a,t_a)  {\,\mathbb{L}}_ \omega \ \!
  \omega(v, t_{ba}).
\label{@}\end{equation}
If we now take into account the evolution equation (\ref{4.2a}), we obtain
from Eq.~(\ref{@LABEL}) the equation
\begin{eqnarray}
\int_0^{\infty} \d v \ \! P_v(x_b,t_b|x_a,t_a) \ \!\partial_{t_{ba}}
\ \!\omega(v,t_{ba}) \ = \
\int \d v\,{P}_v(x_b,t_b|x_a,t_a)  {\,\mathbb{L}}_ \omega \ \!
  \omega(v, t_{ba}).
\label{@}\end{eqnarray}
This result must hold for any distribution $P_v(x_b,t_b|x_a,t_a)$,
thus proving the K-M equation (\ref{4.11aa}) for $\omega(v,t)$.

\vspace{3mm}

\section*{Appendix~B}\label{Ap2}

Here we show how to define uniquely the classical Hamiltonian
$F_{\rm cl}({H}(p,x))$ if the initial Hamiltonian $H(p,x)$ depends
not only on $p$ but also on $x$,
This will be done with the help of
the K-M equations.

We start with an observation that when
the unsmeared K-M operator has no explicit time dependence, i.e., when
\begin{eqnarray}
{\mathbb{L}}_v(-\partial_x, x)\  = \ \sum_{n=1}^{\infty}
(-\partial_x)^n D_v^{(n)}(x)\, ,\label{B.1}
\end{eqnarray}
then the {\em unsmeared} classical Hamiltonian $H_{\rm cl}(p,x)$ has the
form
\begin{eqnarray}
H_{\rm cl}(p,x) \ = \ - \sum_{n=1}^{\infty} (-i p)^n D_v^{(n)}(x)\, ,
\label{B.2}
\end{eqnarray}
provided we define the path integral
by time slicing in the post-point form.
This is done by rewriting
the short-time matrix elements
 $\langle
x_{n}|\exp(-\hat{H}\tau)|x_{n-1}\rangle$
as
integrals
$\int (\d p_n/2\pi) \langle x_{n}|p_n\rangle \langle p_n|\exp(-\hat{H}\tau)|x_{n-1}\rangle$
and
defining the classical
Hamiltonian via the identity
\begin{eqnarray} \!\!\!\! \!\!\!
\langle p_n|\exp(-\hat{H}\tau)|x_{n-1}\rangle \ \equiv \
e^{-\tau H_{\rm cl}(p_n,x_{n-1})}\,{e^{-ip_nx_{n-1}}}
.\label{B.3}
\end{eqnarray}
Then to order ${\mathcal{O}}(\tau^2)$ the
Hamiltonian $H_{\rm cl}(p,x)$ coincides with (\ref{B.2}).
Full discussion of this relation
can be found in Ref. \cite{PI}. The K-M
equation is basically a Schr\"{o}dinger-type equation with the
non-hermitian Hamiltonian $\hat{H}(p,x) =
\sum_{n}(-i\hat{p})^n D_v^{(n)}(\hat{x})=-{\mathbb{L}}_v(-\partial_x, x)$.

Let us now see how the K-M equation looks for the smeared
distribution. This  will allow us to identify the
{\em smeared} classical Hamiltonian $F_{\rm cl}(H)$. To this end we
note that from the definition (\ref{@4.4a0}) the smeared K-M
coefficients can be written as
\begin{eqnarray}
D^{(n)}(x) &=& \frac{1}{n!} \lim_{\tau
\rightarrow 0} \frac{1}{\tau} \int_{0}^{\infty} \d v \ \!
\omega(v,\tau) \int_{-\infty}^{\infty} \d y \ \! (y-x)^n \langle y|
e^{-v \hat{H}\tau} |x \rangle\nonumber\\[2mm] 
&=& \frac{1}{n!} \lim_{\tau
\rightarrow 0} \frac{1}{\tau} \int_{-\infty}^{\infty} \d y \ \!
(y-x)^n \langle y| e^{-F(\hat{H})\tau} |x \rangle .
\end{eqnarray}
Furthermore, we can insert in front of $e^{-F(\hat{H})\tau}$ a
completeness relation $\int \d p \ \! |p\rangle
\langle p|= \hat{{\bf 1}}  $, and make use of the fact that only terms up to order
${\mathcal{O}}(\tau)$ are relevant in $e^{-F(\hat{H})\tau}$. This
leads directly  to
\begin{eqnarray}
\mbox{\hspace{-6mm}}\frac{1}{n!} \lim_{\tau \rightarrow 0} \frac{1}{\tau}
\int_{-\infty}^{\infty} \d p \,\d y \ (y-x)^n \langle y|p \rangle
\langle p| e^{- F(\hat{H})\tau} |x \rangle  \ = \  \frac{1}{n!}
\lim_{\tau \rightarrow 0} \frac{1}{\tau} \int_{-\infty}^{\infty}
\frac{\d p \,\d y}{2\pi} \ y^n e^{ipy} e^{-F_{\rm cl}(H)\tau} \!\, .
\label{B.5}
\end{eqnarray}
Here $F_{\rm cl}(H)$ is obtained from
$F(\hat{H})$ by using the commutator $[\hat p,x]=-i$
to move all $\hat p$'s to the left of all $x$'s
and dropping the hats. Eq.~(\ref{B.5}) can be further simplified if we write
$F_{\rm cl}(H) = \sum_n p^n f_n(x)$ and the $y$-integral as the $n$-th
derivative of $\delta$-function, i.e.
\begin{eqnarray}\nonumber \\[0mm]
D^{(n)}(x) \ &=& \ \frac{(-i)^n}{n!} \lim_{\tau
\rightarrow 0} \frac{1}{\tau} \int_{-\infty}^{\infty}\d p \
 e^{-\tau F_{\rm cl}(H)}\ \! \partial_p^n \delta(p)
\nonumber \\[2mm]
&=& \ \frac{i^n}{n!} \lim_{\tau
\rightarrow 0} \frac{1}{\tau} \left.\partial_p^n \left[e^{-\tau
F_{\rm cl}(H)}\right]\right|_{p=0} = (- i)^n f_n(x)\, .
\end{eqnarray}

\comment{As a second situation we consider here the situation where
the initial ``classical" Hamiltonian $H = K(p) + V(x)$ then the
corresponding ``quantum" Hamiltonian is unambiguously $\hat{H} =
K(\hat{p}) + V(\hat{x})$. From this one obtains straightforwardly
$F(\hat{H})$ and can cast it (unambiguously) in the  $p$-$x$ ordered
form. Applying again the time-sliced regularization of the path
integral one inserts in $\langle \exp(-t_{ba} F(\hat{H}))
\rangle_{ba}$ a complete set of position and momentum eigenstates,
and obtains the product of matrix elements $\langle
p_i|\exp(-F(\hat{H})\epsilon)|x_{i}\rangle$ with plane waves
$\langle x_i|p_{i-1}\rangle$,  with $\epsilon = t_{ba}/N$. By the
Trotter formula the matrix element $\langle
p_i|\exp(-F(\hat{H})\epsilon)|x_{i}\rangle$ can be for large $N$
approximated by $\langle p_i|(1 - F(\hat{H})\epsilon)|x_{i}\rangle$.
If we fix {\em a priory} operator ordering in $F(\hat{H})$ to be
$p$-$x$ ordering the above matrix elements are {\em unambiguous}.
$(F(H))_{cl}$ is then obtained through relation (\ref{B.3}). It is
again obvious that to order ${\mathcal{O}}(\epsilon^2)$ we can
identify $_{\rm cl}F(H) = (F(H))_{xp}$.
 It should, however, be stressed that when a path
integral is regularized differently than via time-slicing, then the
explicit form of $\bar{H}$ will be different (it will be
regularization dependent). It is particulary challenging to find
$\bar{H}$ in the framework of the covariant formulation of path
integrals~\cite{KCH}.} \vspace{3mm}

\section*{References}
%


\begin{thebibliography}{99}
\bibitem{remark}
In the literature,
Eq.~(\ref{5.17}) is
often referred to as the Klein-Gordon
kernel~\cite{GS}. This name is misleading since a
true kernel for the Klein-Gordon equation must include also the
negative energy spectrum which itself reflects an existence of an
independent charge-conjugated solution -- antiparticle.
\end{thebibliography}
\end{document}